# Understanding and addressing the resistance towards autonomous vehicles (AVs)

S. Nordhoff [a]

[a] Department Transport & Planning, Delft University of Technology, The Netherlands

**ABSTRACT**

Autonomous vehicles (AVs) are expected to bring major benefits to transport and society. To exploit this potential, their acceptance by society is a necessary condition. However, AV acceptance is currently at stake. AVs face resistance by bystanders and local communities. This is especially problematic because resistance can prevent the implementation and use of AVs, threating road safety and efficiency. Resistance towards AVs has been largely overlooked by research and practitioners. The present study performed a qualitative and quantitative text analysis of comments submitted by locals in San Francisco (SF) to the California Public Utilities Commission (CPUC) on the fared deployment of AVs in SF. The results of the analysis are synthesized, and a conceptual framework explaining and predicting resistance is proposed. The framework posits that the occurrence of resistance is a direct result of the perception of threats, which is determined by individual and system characteristics, direct and indirect consequences of system use, reactions of others, and external events. Perceived threats pertained to safety, traffic, travel choices, energy consumption and pollution, social equity, economy, and society. AVs as threat to safety was associated with their unpredictable, and illegal driving behavior, as well as producing conflict situations. The lack of explicit communication between AVs and other road users due to the absence of a human driver behind the steering wheel negatively contributed to perceived safety and trust, especially for vulnerable populations in crossing situations. Respondents reported a negative impact on road capacity, congestion, and traffic flow, with AVs blocking other road users, such as emergency vehicles. AVs were conceived as a threat to the transition towards more sustainable mobility with inclusive, multimodal public transit options. Inaccessible vehicle design contributed to the exclusion of vulnerable groups with disabilities. The scientific dialogue on acceptance of AVs needs to shift towards resistance as the 'other' essential element of acceptance to ensure that we live up to our promise of transitioning towards more sustainable mobility that is inclusive, equitable, fair, just, affordable, and available to all.

*Keywords:* Autonomous vehicles; acceptance; resistance; local communities; perception of threats; vulnerable populations

1. **Introduction**

The California Public Utilities Commission (CPUC) approved the fared deployment of autonomous vehicles (AVs) in San Francisco (SF) (CPUC, 2023). As the operation of AVs is limited to specific domains, with these vehicles currently not being able to drive everywhere in all conditions, the present paper defines these vehicles as SAE Level 4 or High Driving Automation (SAE International, 2021). These AVs currently face resistance from locals outside these vehicles, stopping these vehicles by placing objects on them (e.g., cones) or stepping in front of them to test their capabilities and interfere with their operation (Thubron, 2023).

The automated vehicle acceptance literature is skewed towards acceptance, applying technology acceptance models to identify the factors predicting acceptance by drivers (Louw et al., 2021), passengers (Pascale et al., 2021), and other road users (Schrauth, Funk, Maier, & Kraetsch, 2021). Most studies used specific early-adopter populations consisting mostly of males, younger- to middle-aged, and tech-savvy individuals. This implies that important user groups who may benefit the most from AVs, such as the vulnerable populations with special needs, have been excluded from the debate to a large extent until now. The technology acceptance models that were developed for the assessment of automated vehicle acceptance, such as the multi-level model on automated vehicle acceptance (MAVA) (Nordhoff, Kyriakidis, Van Arem, & Happee, 2019), consider resistance only as a side phenomenon without explaining its underlying processes.

The resistance towards AVs is still little understood. In comparison to conventional cars, AVs are mobile, situationally aware, being able to adapt and communicate with their environment (Winfield, 2012), and have higher sensing capabilities than conventional vehicles, creating privacy and security (e.g., hacking) issues (Bloom, Tan, Ramjohn, & Bauer, 2017; Chen, Khalid Khan, Shiwakoti, Stasinopoulos, & Aghabayk, 2023). Current work also revealed concerns related to safety, trust (Chen et al., 2023), affordability, unemployment and financial insecurity (Agrawal et al., 2023). Scholars investigating the acceptance of renewable wind energy have proposed that community or local acceptance is a key determinant for societal acceptability. Resistance is low if the conditions for distributional justice (sharing of costs and benefits), procedural justice (equal opportunities of all relevant stakeholders for participation in the decision-making process), and trust in the information and intentions of the investors and actors outside the community are met (Wüstenhagen, Wolsink, & Bürer, 2007).

Resistance should not be seen as dysfunctional behavior representing an obstacle or barrier to overcome, or that needs to be investigated to improve the uptake and use of AVs (Milakis & Müller, 2021; Van

Wynsberghe & Guimarães Pereira, 2022). Instead, it should be considered as functional orientation that can occur as the result of legitimate concerns associated with a change (Marakas & Hornik, 1996). It should be approached with curiosity rather than stigmatization, being assigned an equal weight in the debate about AVs than acceptance. Neither phenomenon is more important than the other; we need to investigate both ends of the spectrum to the same extent to ensure that the development, design, and deployment of AVs reflect the diverse needs, views, and concerns of all socially relevant individuals and groups in and around AVs. The design of AVs should be a process being open to producing different outcomes representing the results of negotiations among socially relevant groups within different sociocultural and political environments until the design no longer creates problems to any group (Klein & Kleinman, 2002; Milakis & Müller, 2021).

Resistance is defined as a psychological reaction or behavior, representing different types of usage behaviors. It can involve disuse (lack of or no use), low level of use, harmful use (Martinko, Zmud, & Henry, 1996), or misuse (Marakas & Hornik, 1996). It can be passive (e.g., excuses, delay tactics), active (e.g., voicing opposite points of view, asking others to intervene, forming coalitions), or aggressive (e.g., strikes, boycotts, sabotage) (Lapointe & Rivard, 2005). Resistance is also inextricably linked to an object or content that is being resisted (e.g., introduction of new technology) (Jermier, Knights, & Nord, 1994; Lapointe & Rivard, 2005). In line with Marakas and Hornik (1996), this paper posits that acceptance and resistance should be placed on one continuum, with acceptance as measure for use at one end of the continuum, and resistance as measure for disuse at the other end of the continuum (Figure 1).

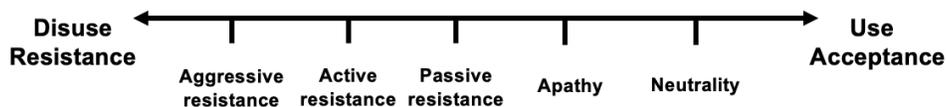

*Figure. 1.* Resistance-acceptance continuum, based on Lapointe and Rivard (2005)

The perception of threats, i.e., expected consequences, is a necessary condition for the occurrence of resistance. In studies examining the resistance towards the implementation of information technology, the perception of threats, a change in the power dynamics between groups with unequal gains, inequity issues, stress and fear, efficacy or outcome expectations represented fertile conditions for the occurrence of resistance. Finally, resistance is linked with initial conditions or subjectivities addressing the differences between individuals or groups of individuals. The interaction of these initial conditions and the object determines the perception of threats, which in turn determine the resistance. Resistance elicits some "triggers", including the actual consequences of system or technology use, events, and reactions of others (e.g., system's advocates, other actors). Individual resistance can change into group resistance (i.e., aggregate of individual resistance behaviors) when individuals' shared perceptions, affect, and responses (i.e., group norms) are activated (Lapointe & Rivard, 2005).

## 1.1. The present study

Resistance can have severe negative consequences, preventing the implementation or use of the system, or making it more difficult for system designers to achieve their objectives (Markus, 1983). An enhanced understanding of the resistance towards AVs as psychological phenomenon is expected to contribute to exploit the benefits of this technology (Shariff, Bonnefon, & Rahwan, 2021). The main objective of this study is to examine the resistance towards AVs, identifying the factors underlying resistance. Comments submitted by the locals of SF to the CPUC on the fared operation of AVs were analyzed using qualitative and quantitative text analysis techniques. Finally, this paper offers a conceptual framework, which synthesizes the results of the data analysis.

## 2. Methodology
## 2.1. Respondents

The present study analysed public comments submitted to the CPUC on the fared deployment of the AVs in SF between February 06, 2020, and August 13, 2023. Next to their comments, respondents provided their name and location of residence.

## 2.2. Data analysis

The data was analyzed in four steps.

First, simple frequency analysis of the most common terms was conducted. In line with Zhou, Kan, Huang, and Silbernagel (2023), the text was preprocessed and cleaned: Part of speech tagging (POS), such as noun, verb, and adjective, was applied to each token (word). Duplicate, and stopwords were removed, and words were transformed to lower cases, and to its root form (lemmatization). Any other noise was also removed, such as characters, digits, hashtags, or hyperlinks. The sentences were tokenized, which means that each sentence was separated into a smaller unit of sentences so that it can be more easily processed by the algorithm.

Second, the main categories and sub-categories were developed using principles of inductive category development form Mayring (2000). The labeling and description of the sub-categories were adjusted by compared the sub-categories with the literature, i.e., studies examining the sub-categories (Kusano et al., 2023; Lehtonen et al., 2022).

Third, a guided Latent Dirichlet Allocation (LDA) model was run to investigate the occurrence of these themes identified in the second step. To identify the occurrences of the sub-themes in the data, seed terms were identified, which were developed inductively from the dataset. It was assumed that each sentence can represent a sub-theme. A sub-theme was assigned a frequency of 1 if at least two seed terms representing a sub-theme were mentioned once in a sentence. The total number of mentions of a sub-theme equals the total number of occurrences of at least two seed terms per sub-theme across all sentences of the data. The analysis was conducted in Python.

Fourth, illustrative comments were selected to portray the meaning of each theme. Multiple mentions of a sub-theme by a respondent were not removed but combined with other mentions of the same sub-theme by the respondent. Consequently, some comments represent collections of sentences mentioned by the same respondents. Each comment was assigned a code, denoted as c, which will be cited as reference for each respondent.

## 3. Results

This section provides the results of the analysis.

### 3.1. Respondents

In total, 325 comments (2283 sentences) were analysed. Most of the respondents seemed to interact with AVs as external road users, while only a small number of respondents reported to have actual experience with these vehicles as passengers. Respondents also represent vulnerable road users with special needs and specific groups (e.g., taxi drivers).

Figure 1 provides an overview of the frequency analysis of the 35 most common terms that were used by respondents.

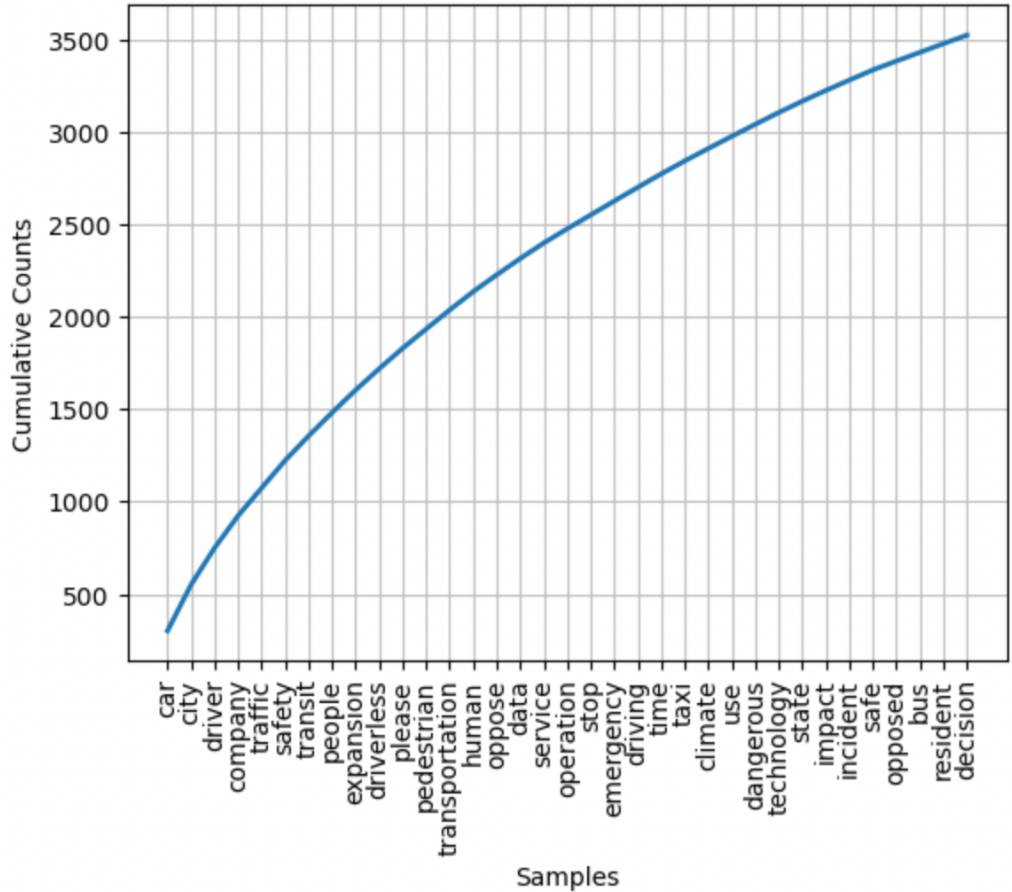

*Figure 1.* Frequency analysis of the 35 most frequently used terms

The qualitative data analysis resulted in the identification of main themes representing the perceived threats associated with the operation of AVs. Table 1 provides an overview of the themes that were identified.

*Table 1.* Overview of data analysis results; main themes and sub-themes (*n*, number of sentences mentioning at least two key words per sentence)

| Main theme | Sub-theme | Meaning | Keywords | *n* |
|---|---|---|---|---|
| Safety | Unpredictability | Unpredictability of vehicle due to erratic, unhuman, unexpected behavior, contributing to low perceived safety and trust. | 'unsafe', 'hack', 'unpredictability', 'unpredictable', 'unexpected', 'erratic', 'unhuman', 'not safe', 'cause accident', 'unsafe', 'crash' | 26 |
| | Illegal driving | AVs engaging in illegal driving, violating traffic rules, e.g., running a red light or stop sign, or parking in bus lane. | 'reckless speeding', 'running red light', 'run red light', 'illegal', 'not safe', 'cause accident', 'crash', 'unsafe' | 0 |
| | Conflict situations | AVs as threat to public safety, being perceived as risky, unsafe, or dangerous, with AVs producing conflict situations by causing collisions, evasive maneuvers by other road users, or an unsafe proximity with other road users. | 'risk', 'public safety menace', 'danger', 'dangerous', 'not safe', 'creeping', 'hit', 'harass', 'struck', 'rear-ended', 'kill', 'cause accident', 'crash', 'unsafe' | 76 |
| | Explicit communication | Lack of explicit communication (e.g., eye contact, hand gestures, verbal communication) with AV due to absence of human driver, contributing to decrease in perceived safety, and sense of control, especially at intersections or in crossing situations, or lack of trust in capability of AV for successful road user detection. | 'unsafe', 'eye contact', 'creeping', 'crosswalk', 'gestures', 'body language', 'cross', 'senior', 'pedestrian', 'cyclist' | 76 |
| | … or AVs as blessing for safety? | Positive safety impacts of AVs, especially for vulnerable populations (e.g., pedestrians, cyclists) given higher driving performance of AVs compared to human drivers due to advanced sensing capabilities of AVs, with AVs obeying traffic rules, and not being prone to human performance decrements and pitfalls while driving. | 'approve', 'support', 'safer than', 'feel safer' | 1 |
| Traffic | Road capacity, congestion, traffic flow | Negative impact on road capacity, congestion, and traffic flow due to perceived larger number of vehicles on road, behaving in unexpected and erratic ways, blocking other road users, such as emergency vehicles. | 'unsafe', 'block', 'nuisance', 'hazard', 'stall', 'add to traffic', 'traffic problem', 'traffic jam', 'nuisance', 'hazard', 'congestion', 'congested', 'delays', 'havoc', 'delay', 'increase congestion', 'disrupt', 'flow', 'ambulance', 'emergency vehicles', 'first responders' | 52 |
| Travel choices | Public transport, walking, cycling use | Negative impact on public transit / transport, and sustainable travel modes, threating transition towards more sustainable mobility with inclusive, multimodal public transportation options. | 'oppose', 'inclusive', 'multi-modal', 'less cars', 'mass transit', 'more sustainable', 'liveable', 'public transit', 'public transportation', 'accessible', 'sustainable', 'liveable', 'not more cars' | 17 |
| Energy consumption and air pollution | Fuel, energy, emissions, pollution | Negative impact of AVs on energy consumption, and air pollution, with AVs running unoccupied, contributing to increase in single vehicle miles travelled (VMT), noise, emissions, pollution, and car dependency, impairing realization of Vision Zero, and transition towards more sustainable mobility. | 'oppose', 'run empty', 'VMT', 'unoccupied', 'Vehicle Miles Traveled', 'more emissions', 'pollution', 'climate crisis', 'environment', 'pollutants', 'space', 'boxes', 'climate change', 'climate emergency' | 21 |
| Social equity | Vehicle design | Insufficient vehicle accessibility, contributing to social exclusion of vulnerable populations with disabilities, pertaining to ordering of vehicle via smartphone app, entering and exiting vehicle, getting buckled on, and receiving assistance with carrying passengers' items to apartment. | 'unsafe', 'curb', 'buckled', 'blind', 'senior', 'accessibility', 'accessible', 'accessibility', 'disabilities', 'accessible', 'disability' | 51 |
| | Public engagement | Insufficient public engagement in decision of trialing and operating AVs on public roads, manifesting perceived sense of injustice and social inequality between residents in SF, particularly traditionally marginalized populations, and representatives of tech companies with unequal distribution of benefits and risks. | 'oppose', 'without consent', 'local participation', 'democratic engagement', 'guinea pigs', 'experimented', 'rich', 'lab', 'testing', 'profits', 'tech companies', 'corporate', 'rich', 'billionaires', 'corporations' | 25 |
| Regulation | Liability | Unresolved liability questions pertaining to holding AVs accountable for traffic violations or accidents. | 'oppose', 'not accountable', 'no accountability', 'account', 'can't be cited', 'above the law', 'violations', 'unaccountable', 'liability', 'liable' | 16 |
| | Transparency | Insufficient transparency pertaining to sharing AV-performance-oriented data. | 'oppose', 'no transparency', 'not transparent', 'lack of transparency', 'share data' | 0 |
| Economy | Unemployment | Expected unemployment among professional drivers due to AVs replacing drivers, and associated fear of consequences due to insufficient governmental | 'oppose', 'lost wages', 'livelihoods', 'take away', 'unemploy', 'unemployment', 'social inequality', 'labor rights', 'replace', 'poverty', 'jobs', 'displayed', 'labor', | 19 |

| | | support assisting drivers in transition. | 'exploit', 'eliminate', 'threat', 'taxi drivers', 'dispensable' | |
|---|---|---|---|---|
| Society | Data privacy | Data privacy concerns pertaining to AV sensors (i.e., laser, radar, camera, lidar) capturing road user data, including video and audio data, engendering civil rights of, e.g., people seeking abortions. | 'oppose', 'surveillance', 'data privacy', 'liberties', 'surveil', 'audio data', 'video data', 'cameras', 'armies', 'patrolling', 'prosecute', 'civil rights', 'dystopian', 'liberties', 'abortions' | 42 |
| | Humanity | AVs as embodiment of artificial intelligence (AI) representing threat to humanity. | 'oppose', 'humanity', 'artificial intelligence', 'threat to humanity' | 0 |

### 3.2. Safety

#### 3.2.1. Unpredictability

This sub-theme addressed the unpredictability of the vehicle, which was associated with its erratic, unhuman, and unexpected behavior, contributing to low perceived safety and trust.

> *"Strongly opposed. AVs are a significant threat to my safety. Their odd behavior (sudden, persistent stops) makes other vehicles move more erratically."* (c94134)

> *"I personally have experienced their unexpected, reasonless stops in the middle of the roadway and intersections. I have observed them start and stop repeatedly after stopping far behind a limit line, and not proceeding."* (c94118)

> *"I've seen self-driving cars make unexpected turns, ride too close to bikes, and speed recklessly. I will never feel safe next to a machine-operated machine."* (c94102)

> *"These AVs have displayed erratic maneuvers across San Francisco that are not in line with how a human driver would typically behave. This unpredictability poses a significant safety hazard to pedestrians, cyclists, and human drivers who share the roads with these autonomous vehicles. As the technology is still relatively new, human drivers are finding it challenging to* *anticipate how these robot cars will respond, given their unique driving patterns."* (c94121)

*"I am so tired of these autonomous vehicles driving around our neighborhood all day and all night. I have witnessed them make so many bad decisions that I simply cannot trust them."* (c94117)

### 3.2.2. Illegal driving

Illegal driving behavior of AVs, violating traffic rules, such as running a red light or stop light, as well as parking in the bus lane, were also reported.

*"I see driverless cars and have found them to often be erratic, unpredictable, and at times in violation of traffic laws."* (c94118)

*"When the light turned green, it proceeded to go around the firetruck to the right and PARK IN THE BUS STOP! Why? It had a perfectly clear road ahead of it. It is illegal for cars to park in the bus stop! If a robocar cannot assess a street situation like this, it should not be on the road at all! Stop the robocars now!"* (c94109)

"*There have been hundreds of documented examples of AVs behaving badly. Cruise even posted a photo of one of their cars running a red light on their 2022 impact report!!"* (c94114)

*"I encountered a Cruise car that ran a stop sign."* (c94115)

*"They can't even follow the rules of the road. I constantly see them signal one direction then proceed in another, or outright fail to yield to pedestrians in crosswalks."* (c94117)

### 3.2.3. Conflict situations

AVs were considered a threat to public safety, being perceived as risky, unsafe, or dangerous, producing conflict situations by causing collisions, evasive maneuvers by other road users, or an unsafe proximity with other road users.

*"Autonomous cars are disrespectful of the elderly, children, parents with children, and everyone else crossing the street. They come to a complete stop at controlled intersections, but proceed through the intersection even when someone is already crossing the street."* (c94118)

*"I have narrowly avoided having the car in which I was riding avoid being struck when the AV turned into the wrong lane."* (c94118)

*"AVs harass pedestrians in marked crosswalks by creeping up on us. I have seen AVs in my neighborhood making right turns without slowing down. This is very dangerous, especially in neighborhoods with lots of kids like mine."* (c94134)

*"Cruise vehicles can't figure out crosswalks and have nearly hit my family."* (c94103)

*"I have heard these robocars are unsafe. I have heard they have killed some small dogs in SF due to not perceiving them. What will happen if a toddler, around the same size as a dog, jumps into the road? Human drivers can respond, but robocars look like they can't. We have heard about a lot of accidents they have caused."* (c94109)

*"I live in the Mission District and have almost been hit twice now by Cruise driverless vehicles while using the crosswalk by my house."* (c94110)

*"Strongly opposed. I've had to jump out of the way of these cars twice."* (c94114)

*"I was taking a walk at night with my partner and an AV did not slow down when we were crossing in front of it. We moved quickly out of the way to avoid getting hit. This threat to safety is unacceptable."* (c94118)

*"While out walking today, two cars failed to give me the right of way when I was in a crosswalk or about to enter one. The Waymo began to make a right turn onto the street I was crossing before I had stepped onto the far curb and was still in the crosswalk. The Cruise vehicle left its stop sign on the far side of the intersection as I was stepping off the curb into the crosswalk. At least with a human driver there is the possibility of establishing eye contact."* (c94118)

### 3.2.4. … or AVs as blessing for safety?

This sub-theme addressed the perceived positive safety impacts of AVs, especially for vulnerable populations (e.g., pedestrians, cyclists), given the better driving performance of AVs in comparison to human drivers, and the advanced sensing capabilities of AVs, with AVs obeying the traffic rules, and not being prone to human performance decrements and pitfalls while driving (e.g., distracted driving). Respondents mentioned the need to apply the same or similar safety standards for AVs currently being applied for human-controlled vehicles.

*"As a frequent pedestrian I've noticed AVs obey traffic rules and respect pedestrians much better than human drivers. They often speed, drive on sidewalks, drive into oncoming traffic, ignore pedestrians in crosswalks, run red lights, etc because of a lack of patience. AVs have unlimited patience."* (c94131)

*"I have taken over a dozen autonomous rides and found them to be safe and effective. They obey the rules of the road and allow people like myself to safely cross the road without worrying about distracted and impatient drivers."* (c94117)

*"As an SF resident that loves biking, I feel much safer around driverless vehicles given their inherently increased visibility and awareness of their surroundings. The data seem to strongly support this; the autonomous vehicles have significantly less incidents per mile driven."* (c94117)

*"I have been hit multiple times by vehicles on the streets and I feel much safer cycling around these driverless vehicles than around humans. They treat every*

*stop sign like a real stop sign (no rolling stops), never speed, and provide wide berth when passing me whether it is on my bike or when I'm just walking normally."* (c94107)

*"As a pedestrian & driver myself, I feel SIGNIFICANTLY safer around these vehicles than the people driving in SF. I watched two human drivers blow through stop signs and nearly run over crossing children just this past weekend."* (c94114)

*"As an avid cyclist, who's twice been hit by cars on SF streets, I very much SUPPORT driverless cars on our streets. I've biked by many, and (unlike many of the drivers in this city) they are always paying attention. I look forward to safer streets with more bikes and less parked cars, and fully believe that AVs can help us get there."* (c94158)

*"I am concerned with the public discourse. Many of the negative comments I've seen exaggerate (or lie about) the problems with these services, cite unproven anecdotes and/or fail to compare their safety record with human driven automobiles. It makes no sense to hold AVs to much higher standards than we currently hold human drivers. The data clearly shows that AVs are much safer than human-driven cars, regardless of any random person's opinion on the matter."* (c94103)

*"I was skeptical about these Cylon death machines until I rode in one. At that point I was an instant fan. Much safer than Uber drivers. Slower. Obeys the signs. Respects cyclists. Isn't pushy or aggressive. They're like big orange cows. Since my first trip, I've taken about fifty trips."* (c94117)

### 3.2.5. Explicit communication

This sub-theme addressed the lack of explicit communication (e.g., eye contact, hand gestures, verbal communication) between AVs and other road users due to the absence of a human driver behind the steering wheel. This lack of explicit communication contributed to a decrease in perceived safety, and lack of perceived sense of control, especially at intersections or in crossing situations, and lack of trust in the capability of the AV to successfully detect vulnerable road users.

*"Human drivers are dangerous but at least there's someone present I can communicate with, yell at to stop. A driverless car will just hit me and keep going and there's nothing anyone can do."* (c94103)

*"I am a blind individual who works in SF. I am concerned that a poorly controlled autonomous vehicle may collide with me and my guide dog while I cross a street."* (c94577)

*"As a cyclist, I have stopped at intersections and waited to make eye contact with drivers who have also stopped, to acknowledge who goes next. This is impossible with a driverless car, as THERE ARE NO EYES."* (c94117)

*"I see driverless cars and have found them without any means to give direct feedback. Human drivers are often this, but I can yell at a human driver to get their attention if needed."* (c94118)

*"You can't communicate with an AV car. It's like watching a newborn drive but at least a newborn can look at you. No thank you!"* (c94110)

*"As a cyclist, I am strongly opposed to autonomous vehicles full stop. Bikes rely on human signals to safely move through car-centered streets – gestures, making eye contact, body language."* (c94102)

*"I am a senior bike rider and I rely on making eye contact with drivers for safety, which can't be done with these cars. It is frightening to come to an intersection with one of these and not know if the "algorhythm" has been tweaked or how it will respond."* (c94121)

*"Even crossing the street in front of an autonomous vehicle is a challenging proposition. There is no one to make eye contact with to give the pedestrian assurance that they are seen and it is safe to cross the street."* (c94110)

> *"As a pedestrian, intersections are terribly dangerous. The only way I know if I'm relatively safe to cross is if I can make eye contact with the driver at the intersection to know that they acknowledge my right-of-way to cross. Driverless vehicles CAN'T make eye contact and therefore make it impossible for me to cross safely. How will I ever know if it is safe to cross? What are the programming contingencies for these deadly machines? What are the actual statistics for how they behave in what circumstances?"* (c94110)

> *"I was so mad, and there was no one to make eye contact with, which is one of the creepiest things."* (c94115)

> *"You can't make eye contact with a robot car to make sure they see you before you cross the street in a crosswalk at a stop sign or stoplight. I don't want to play chicken with a robot car not knowing whether they'll see me in the street."* (c94110)

### 3.3. Traffic implications
### 3.3.1. Road capacity, congestion, traffic flow

A negative impact on road capacity, congestion, and traffic flow was mentioned due to the expected larger number of vehicles on the road behaving in an unexpected and erratic way, e.g., stopping unexpectedly, impairing merging or splitting situations, or blocking other road users, such as emergency vehicles.

> *"I am strongly against "self-driving" cars. I've witnessed these cars stop in the middle lane of Hyde Street on a green light as ambulances are trying to pass."* (c94109)

> *"It is DEEPLY troubling and concerning that driverless cars have been involved in these and numerous other incidents, including getting in the path of first responders at numerous scenes of crisis, including a mass shooting."* (c94110)

> *"The self-driving vehicles are worsening congestion. I saw one hit a car with an elderly, minimal English-speaking driver who had no recourse to follow up, and then it blocked a bus stop."* (c94115)

*"I've lost counts of the number of times I've seen them block intersections, bike lanes, buses, street cars ... you name it!"* (c94114)

*"I witnessed a robocar pull up behind a firetruck. The robocar stopped in the middle of the intersection, blocking northbound traffic. All the cars were honking, not realizing it was a robocar. Many of us gathered to watch. I clocked it - it took 10 minutes before the robocar backed up out of the intersection."* (c94109)

*"Oppose; Cruise and Waymo AVs are causing havoc in our streets. Multiple times a day they stop randomly and block traffic. Often they interfere with both transit and emergency services."* (c94117)

*"These vehicles roam around in a pack and make it hard to get away from or merge into. It was fun to see them once in a while at first, but I am now strongly against the expansion of the program due to the sheer annoyance and frustration that they are jamming up our streets."* (c94118)

*"I am vehemently opposed to any expansion of AVs on San Francisco streets. More than once, I have been on a bus carrying 20+ people to their destinations that could not move forward because an AV carrying no one was immobile in front of it for no apparent reason."* (c94134)

*"They have greatly increased congestion already. When they encounter an un-programmed situation, they simply stop, blocking everyone behind them."* (c94122)

*"I witnessed one of these driverless vehicles. It blocked the 44 Muni bus which was unable to go around the car. The Muni driver opened the doors to let off passengers, including elderly and disabled folks, because the bus clearly was not going anywhere anytime soon. After waiting 20 minutes with no resolution in sight, we left with the stopped vehicle still in the middle of the road and the 44 stuck behind."* (c94118)

> *"I have personally witnessed these vehicles disrupt the flow of traffic for no apparent reason on more occasions than I can recall. Most commonly, I have seen Cruise vehicles stopped in the middle of the street or in front of green lights, despite there being no other vehicles, objects, or pedestrians in their path."* (c94121)

### 3.4. Travel choices implications

#### 3.4.1. Public transport, walking, cycling use

The negative impact of the operation of AVs on the use of public transport, and sustainable travel modes was mentioned due to traveler's increased use of AVs, and an expected redirection of investments from public transit to AVs. Respondents expected a negative impact on the affordability of public transport, threatening the transition towards more sustainable mobility with inclusive, multimodal public transportation options.

> *"I want to see less money invested in self driving cars and more towards public transit, making transit accessible affordable and even free. We have an active bus metro and subway service in the city that needs much more support and attention to make accessible and safer for residents to use."* (c94112)

> *"I strongly urge you to consider very seriously why we are allowing these AVs to take over our streets instead of investing in TRULY sustainable modes of transportation, like bicycling, walking, and accessible, affordable public transit."* (c94109)

> *"The money being spent on these vehicles must be redirected toward public transit and active transit infrastructure that San Francisco badly needs. Buses are packed, bike lanes are a mess, yet we are here talking about driverless drains on our resources."* (c94112)

> *"I am opposed to this. This forces us to build cities car sized instead of people sized, making it inherently unwalkable."* (c94122)

> *"They prevent development of SF into a city with multi-modal transportations favoring all people."* (c94115)

*"We don't want AV cars on our streets. We want less cars, more sane bike lanes, and more public transportation."* (c94110)

*"We have concerns about the affordability and that there have been no requirements set that AV fared passenger service would be affordable. People in our community are often the most low-income and rely on public transit. We have concerns that AVs will erode the public transportation system and we will see bus and subway service reduced in areas that AVs are servicing."* (c94122)

*"San Francisco needs to focus on increasing pedestrian spaces, and on improving public transit. Keep self-driving cars OUT of San Francisco. Car centric eco structure kills. We need more road closures (like JFK), more bike/skate paths, and better busses."* (cD09C9)

### 3.5. Energy consumption and air pollution
### 3.5.1. Fuel, energy, emissions, pollution

This sub-theme covered the negative impact of AVs on energy consumption and air pollution, with respondents reporting incidences of AVs running unoccupied, contributing to an increase in single vehicle miles travelled (VMT), noise, emissions, pollution, and car dependency, impeding the realization of Vision Zero and the transition towards more sustainable mobility.

*"Most self-driving cars I see are empty. This is just releasing more emissions into our already damaged environment."* (c94109)

*"The last thing we need is more cars of any kind. We are in a climate emergency. We should be reducing vehicle miles traveled (even of electric vehicles)."* (c94131)

*"They are incredibly wasteful. Climate change is wreaking havoc. Invest in real infrastructure that we already know works."* (c94103)

> *"No more driverless cars, please. Our world, our city, needs to move away from cars in order to have a livable planet for future generations. I fail to see how expanding the use of driverless cars driving around San Francisco completely empty is a step toward a sustainable future."* (c94112)

> *"AVs do not actually solve any problem related to the climate crisis – they just exacerbate the problems by adding more VMT to San Francisco, at a time when the climate is warming seemingly exponentially. AVs are not the answer."* (c94114)

> *"I have seen these cars sitting idly blocking traffic, creating more pollution for everyone. Most of the time I see these cars COMPLETELY EMPTY. Simply taking up space in a congested city, no proving service to anyone at all. I love city life for the people, not empty trash boxes taking up space."* (c94110)

> *"Finally, we are in a climate crisis (remember wild fires?). We should been encouraging the increased use of public transportation, biking, and walking instead of flooding the streets with more cars."* (c94110)

### 3.6. Social equity
### 3.6.1. Vehicle design

This sub-theme addressed the lack of vehicle accessibility, contributing to the social exclusion of vulnerable populations with disabilities. Accessibility pertained to ordering the vehicle via smartphone app, entering the vehicle, getting buckled on, exiting the vehicle, and assisting passengers with carrying items to the apartment.

> *"I've heard that autonomous vehicles pick up and drop off passengers in the middle of the street, not at the curb. Making blind passengers go out into traffic to find an autonomous vehicle is extremely dangerous. I need assurances that the AV companies have made their smartphone apps fully accessible for blind passengers and that a reliable method for blind passengers to locate vehicles has been proven to be safe and effective."* (c94577)

> *"They are not even accessible. They never pull up fully to the curb nor do they have any drivers to help folks get in the car."* (c94117)

> *"As a senior, I could not use one because there would no one to assist me into and out of the car and ensure that I was safely buckled in."* (c94118)

> *"ZERO ACCESS TO PERSONS WITH DISABILITIES. They completely violate federal law, specifically the Americans with Disabilities Act. They have been designed from the ground up to exclude people with disabilities."* (c94114)

### 3.6.2. Public engagement

Insufficient local engagement in the decision of trialing and operating AVs on public roads was reported, manifesting a perceived sense of injustice and social inequality between locals of SF and representatives of the tech companies, and unequal distribution of benefits and risks.

> *"These vehicles are being foisted on the people of SF without their consent for the purpose of facilitating a small number of companies in their quest to establish fleets of vehicles under their complete control to profit from. I hope the CPUC will enact more local participation and democratic engagement."* (c94131)

> *"When giant corporations get to use San Francisco as a laboratory (without our consent) we effectively become their lab rats. Further expanding the scope of these programs will NOT help the city."* (c94122)

> *"These companies are using us all as guinea pigs to test their highly experimental software. It's deeply unacceptable that these cars were ever allowed onto our streets without our permission."* (c94110)

> *"We did not consent to this. We are fed up with being experimented on by corporations."* (c94117)

*"They are yet another technological pipe dream that will make few companies rich while using real life San Franciscans as guinea pigs."* (c94115)

*"I do not appreciate having my life and limb at risk as an unwilling participant in their beta testing on our public streets."* (c94117)

*"AV companies are putting San Franciscans at risk for private profit. It's time to prioritize public safety (and public transit!)."* (c94134)

*"Should they ever get it to a safe and useable state (something I am deeply skeptical is even possible), it will be thanks to the thousands of San Franciscans who were literally put into harm's way to do so. Yet those San Franciscans will receive zero compensation – those profits will stay with the car companies."* (c94114)

*"Stop these vehicles, it's a gimmick, a dizzyland ride to make a few rich, does it benefit society, no!"* (c94110)

### 3.7. Governance
### 3.7.1. Liability

This sub-theme addressed the lack of liability of the AV companies in terms of holding their vehicles accountable in case of traffic violations or accidents caused by the AVs.

*"Who will bear the responsibility, when a driverless vehicle breaks the law or causes injury or death to someone in our city?"* (c94102)

*"A driverless car will just hit me and keep going and there's nothing anyone can do, including holding the company accountable."* (c94103)

*"If there is an incident on the road, who do you exchange phone numbers with? Where is the accountability?"* (c94118)

*"The companies are not acting in good faith because you are not holding them to account. They cannot be cited."* (c94118)

*"Robotaxis are effectively above the law. Their fleets cannot be cited for traffic violations."* (c94110)

*"My first gripe with Cruise / Waymo specifically is that they have no accountability to any city or county of San Francisco authority."* (c94107)

*"There is no punishment for an AV killing someone. An innocent civilian death is just accepted with no real consequences. At least a drunk driver will be charged with a crime. When someone has an accident, there are consequences. DL suspension, fines, jail, just dying in the crash. None of those exist for an AV."* (c95073)

*"In car-on-person accidents, there is more at stake than just financial liability; there is also the possibility of criminal charges. When driverless cars commit a crime, who is charged with the crime?"* (c94112)

### 3.7.2. Transparency

Insufficient transparency was mentioned with regards to AV companies sharing performance-oriented AV data, with respondents calling for an independent assessment of the impacts of AVs on transport and society.

*"I demand that AV companies share unredacted incident data with the public and city agencies. The lack of transparency is alarming. By withholding incident data, AV companies are evading responsibility, forcing the public and city agencies to rely on social media posts to understand the extent of the problems they cause. We need a robust and independent reporting system to address these concerns effectively."* (c94110)

*"As they refuse to share incident data, the public, as well as city agencies, must rely on social media posts to determine the extent of the problems they cause. A robust and independent reporting system must be put in place."* (c94110)

*"If these cars are to operate on our streets, they should have to share all safety data with the city, like SFMTA does for our bus fleet."* (c94110)

*"While Cruise may claim to have a clean safety record in terms of fatalities, their reporting provides zero data on the frequency of glitches/adverse events that disrupt the flow of. Until this data is available and addressed, these companies should not be permitted to operate vehicles without a backup driver."* (c94121)

*"I live in the Richmond district and recently there has been a huge uptick in AV with zero insight into the legal and safety ramifications of having these cars on the street."* (c94118)

### 3.8. Economy
### 3.8.1. Unemployment
This sub-theme addressed the expected unemployment among professional drivers due to their replacement by AVs, and associated fear due to insufficient governmental support assisting drivers in this transition.

*"Cruise and Waymo are a disaster for workers and safety. Please stop them. This is a labor and safety issue that we all need to be concerned about as AI makes more and more of us dispensable."* (c94102)

*"Moving humans from behind the wheel of a car to even more invisible positions in call centers & support cars makes them even easier to exploit and quashes unionization efforts. A just transition away from cars and car dominance means taking care of workers who are most affected by it, not just relegating them to more exploitable roles."* (c94122)

*"Not only that, but they are taking away jobs! This is a major concern. Our government has no plans to help all the people who are about to be displaced by technology. These are a horrible idea and should be eliminated."* (c94110)

*"This expansion that will remove more jobs is yet another nail in the coffin of this once great city. Whose job will be replaced next? Quoting from a line in an old union song, "Which side are you on? Which side are you on?*

>*People or Machines?*
>*Human or Artificial Intelligence?*
>*Working People or Corporations?*
>*Human Autonomy or The Surveillance State?"* (c94110)

*"The crucial service of taxi drivers has been severely cut (with people losing their livelihoods) because of Uber and Lyft, and now, Waymo and Cruise is heading down that same destructive path."* (c94110)

*"These driverless vehicles threaten the livelihoods of taxi drivers, many of whom have not recovered from the pandemic and the explosion of unregulated Uber and Lyft vehicles in SF."* (c94102)

*"How about replacing musicians, sportsmen, classroom teachers, restaurant waiters, court judges and all other human roles with robots? Do you want to support this vision?"* (c94597)

### 3.9. Society
#### 3.9.1. Data privacy

Data privacy concerns were mentioned with the AV sensors (i.e., laser, radar, camera, lidar) constantly capturing data from road users around AVs, including video and audio data, engendering civil rights, of e.g., people seeking abortions.

*"These tech companies have purchased the ability to collect information uninhibited and surveil the people of SF. Automation can be a useful tool to help the working class when regulated appropriately, but the AVs using San Francisco's streets for beta testing are an overreach by wealthy technocrats that see our privacy & data as an unexploited resource for their taking. Knock it off."* (c94131)

> *"The dystopian potential these machines present is frightening. Gone are the days when you could walk through San Francisco without being constantly scanned and analyzed by autonomous camera turrets."* (c94122)

> *"It makes me very uneasy to have dozens of driverless vehicles sent out by tech companies equipped with cameras recording and using images for who knows what. There is no way to escape them if one does not wish to consent to their image being recorded and exploited."* (c94112)

> *"These vehicles are constantly capturing video and audio data of their surroundings, which puts us under constant surveillance. Cruise and Waymo have thus installed a surveillance state. Get these surveillance machines off our streets."* (c94117)

> *"These vehicles are equipped with cameras that are used by Cruise and Waymo for both their own internal purposes and the data are shared with law enforcement. We do not need armies of privately owned security cameras patrolling the streets of San Francisco."* (c94103)

> *"Surveillance: This unprecedented invasion of the public's privacy will likely have far-reaching effects on the rights of the general public. The Sacramento police department has forwarded surveillance data to states which could prosecute those seeking an abortion. A city-wide, moving network observing and analyzing everything that happens outdoors is something out of a dystopian movie, not a democratic society."* (c94110)

### 3.9.2. Threat to humanity

Respondents expected a negative impact of AVs on humanity, with AVs as embodiment of artificial intelligence (AI) representing a threat humanity in general.

> *"At a time when many prominent scientists are raising a red flag regarding the dangers of AI to humanity, why would you promote these Autonomous vehicles? These vehicles are anti-human and should be banned. Do you want* *to live in Robot City? For the sake of humanity please do not approve the expansion of Waymo and Cruise operations. Thank you."* (c94110)

*"As robot taxis they cannot help travelers lift luggage, shoppers lift groceries, disabled folks lift walkers or wheel-chairs into vehicles."* (c94122)

*"As a nurse, I'm painfully aware of the need of people with disabilities to have assistance to climb steps, to have help with packages, and to have help when someone is ill and needs medical care."* (c94110)

*"I do not see how these autonomous vehicles, without drivers, are an improvement for the elderly and disabled over regular cabs -- operated by friendly, helpful cab drivers who can help people with doors, luggage, and even carrying groceries upstairs and into homes."* (c94110)

*"We humans need to preserve and foster social interaction, for our balance and sanity. The use of robot taxis is yet another step towards a lifestyle that goes against natural human instincts and, as such, is an attack on the human spirit. Our leaders must defend the human spirit. The world needs more humanism, not less."* (c94597)

### 3.10. Synthesis: Conceptual framework

The results of the analysis are synthesized in a conceptual framework (Figure 2). The framework posits that the occurrence of resistance is a direct result of the perception of threats, which is a function of individual and system characteristics, direct and indirect consequences of system use, reactions of others, and external events. Individual characteristics include, but are not limited to, the individual's age, disability, vulnerability, and knowledge. System characteristics include, but are not limited to, lack of explicit communication, vehicle unpredictability, and vehicle design. Direct consequences of use include, but are not limited to, the experience (as passenger), and as road user experiencing AVs in conflict situations. Reactions of others include, but are not limited to, word-of-mouth, public engagement by decision-makers and other stakeholders, and governance and regulation. External events include, but are not limited to, public engagement by decision-makers and other stakeholders, and governance and regulation.

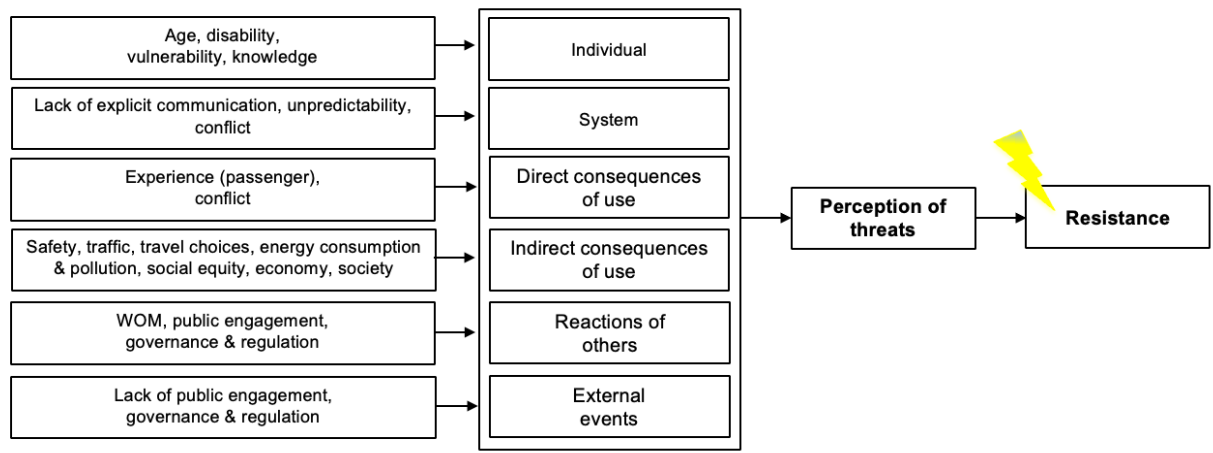

*Figure 2.* Conceptual model explaining and predicting resistance

## 4. Discussion

This study reports the results of the analysis of public comments submitted to the CPUC on the fared deployment of AVs in SF. The data analysis resulted in the extraction of main themes representing the perception of threats associated with the operation of AVs. A conceptual framework synthesizing the results of the analysis is proposed, which explains and predicts the occurrence of resistance towards AVs. The framework proposes that the occurrence of resistance is the direct result of the perception of threats, which are determined by individual and system characteristics, direct and indirect consequences of system use, reactions of others, and external events. The literature on end-user acceptance of AVs has emphasized the relevance of utility-based domain-specific and emotional symbolic-affective factors, and a limited role for moral-normative factors (such as perceived risks) (Nordhoff et al., 2019). In contrast, our study has shown that bystander or local community acceptance may be influenced to a large extent by the perceived risks or threats associated with the operation of AVs. Our analysis has also revealed that direct experience with AVs as passengers might alleviate the perceived threats, which is in line with research studies documenting a positive effect of experience or familiarity on attitudes and acceptance (Penmetsa, Adanu, Wood, Wang, & Jones, 2019; Xing, Zhou, Han, Zhang, & Lu, 2022).

The indirect consequences of system use capture the longer-term impacts of AVs on safety, traffic, travel choices, energy consumption and pollution, social equity, the economy and society, with respondents reporting a perceived negative effect of AVs on these aspects. Studies have shown that vehicle automation is expected to have a positive impact on safety, travel time, highway and intersection capacity, fuel efficiency, and emissions, and a negative effect on vehicle miles travelled, while the effects on energy consumption, social equity, the economy, and society are largely unknown (Milakis, van Arem, & van Wee, 2015). Particularly, the fear of unemployment is a common and legitimate concern raised by the public, which should be addressed by implementing processes, procedures, cultures, and values ensuring ethical behavior in supporting workers in the transition (Winfield & Jirotka, 2018).

Most of the studies assessing the safety impacts of AVs were conducted in simulated rather than real traffic environments using vehicle data. Therefore, the safety implications of AVs remain unclear (Tafidis, Farah, Brijs, & Pirdavani, 2022). Data from naturalistic driving studies with and without a trained operator behind the steering wheel of a AV vehicle has shown that contact events between AVs and other road users did not result in severe or life-threatening injuries, with the AVs being more capable of avoiding collisions compared to human drivers, and that collisions resulted from the interactions with human drivers (Schwall, Daniel, Victor, Favaro, & Hohnhold, 2020). Our study has shown that AVs were perceived as threat to public safety given their unpredictable, erratic, and unexpected behavior, violating traffic rules, and causing conflict situations with other road users. Conflict situations included AVs causing accidents, eliciting evasive maneuvers by road users, or creating an unsafe proximity with road users (Kusano et al., 2023). The discussion of what constitutes acceptable safety as embodied by questions, such as 'How safe is safe enough?', 'Safe enough for what?', or 'Safe enough for whom?' is ongoing (Cohen et al., 2020; Liu, Yang, & Xu, 2019; Shariff et al., 2021; Stilgoe, 2021), and will perhaps not be closed anytime soon. The expected positive safety benefits are more likely to be achieved with an increase in the level of automation, cooperation, and penetration rate (Milakis et al., 2015). It has been shown that the public may have unrealistic expectations about the safety of automated vehicles, expecting higher levels of safety from an automated than from a human driven vehicle (Shariff et al., 2021).

This research has also reported a perceived negative impact of the operation of AVs on road capacity, congestion, and traffic flow, with AVs blocking other traffic, such as emergency vehicles. It has been shown that 82% of first responders did not receive AV-related safety training, and 41% of respondents had little knowledge about AVs, and 44% did not trust AVs (Liu et al., 2023). Particularly the interaction between AVs and emergency vehicles may represent a highly emotional topic for the public, which can have serious life-and-death implications for drivers of the emergency vehicles, other traffic, or people losing a person due to AVs impairing the efficient operation of first responders. Ensuring the safe and efficient interaction between emergency vehicles and AVs could thus be a main contributing factor to promote public acceptance or reduce resistance. External communication devices supporting explicit communication between AVs and external road users were not designed from the perspective of first responders. Similar to treating cyclists as distinct user group with unique communication needs (Berge, Hagenzieker, Farah, & de Winter, 2022), it can be argued that first responders should be treated as specific road user group requiring tailored AV communication. We recommend future research to examine to what extent external Human Machine Interfaces (eHMIs) as explicit communication can support first responders in their interaction with AVs, and determine the specific design characteristics positively impacting safety, efficiency, and acceptance.

This research has also shown that data privacy concerns and its wider societal implications explained resistance, with AVs constantly capturing audio and video data of road users. The data privacy concerns are legitimate, encompassing the collection of demographic information (e.g., driver's license, real-time location, travel behavior), and non-verbal communication (e.g., body movements) (Khan, Shiwakoti, Stasinopoulos, & Warren, 2023). It was revealed that data privacy concerns were a delimiting factor for the acceptance and use of AVs (Zmud, Sener, & Wagner, 2016). Bloom et al. (2017) revealed that respondents' discomfort was highest for the most privacy invasive scenarios involving AVs (vehicle tracking), and lowest for the least privacy invasive scenarios (image capture). To alleviate concerns, locals could be educated about the potential benefits of the large-scale data collection and analysis (e.g., finding of Silver Alert citizens) (Bloom et al., 2017), and be given the possibility to 'opt out' of the analysis of their data. Respondents also mentioned the lack of democratic participation and engagement in providing consent for trialing AVs on public roads, manifesting a perceived sense of injustice and social inequality between citizens and representatives of the tech companies with an unequal distribution of benefits and risks. Future research should examine to what extent engagement and participation (e.g., war gaming methodology, citizen juries) can alleviate concerns of the locals, and promote understanding and knowledge through negotiation and compromise (Birhane et al., 2022; Fraade-Blanar & Weast, 2023).

Insufficient liability and transparency of the AV companies in terms of holding their vehicles to account, and publicly sharing data was mentioned as other factors underlying resistance. Clarifying the responsibilities and roles of stakeholders involved in the deployment of AVs, especially in the case of accidents, having legislation in place, and promoting transparency about the data collection by AVs, and incidences involving AVs could mitigate the occurrence of resistance It remains to be seen to what extent the resistance towards AVs changes when AVs cater for real-world needs and desires, including locals in the development, design, and deployment of these vehicles. For example, the current lack of accessibility to these AVs by vulnerable populations could be directly addressed by the AV companies.

### 4.1. Limitations and implications for future research

First, the data represents the subjective perceptions of respondents. Future research should perform longer-term ethnographic studies with bystanders and local communities in areas with AVs to learn about their lived experiences and interactions, and differences in resistance across situations, and between individuals and groups, making observations, and collecting self-reported data from interviews, focus groups, and surveys, and behavioral and physiological data collected from sensors deployed on respondents.

Second, the comments that were subjected to the present analysis were publicly available, meaning that later comments may have been influenced by previous comments. Future research is needed to assess the extent to which the themes identified in this study can be generalized to other 'bystander groups' and local communities.

Third, the calculation of the occurrence of the sub-theme was based on the co-occurrence of at least two seed terms per sentence, which explains the low number of sub-theme occurrences in some instances. Future research should exploit this technique, including the wider semantic context in neighboring sentences.

Fourth, socio-demographic information of the respondents posting the comments were missing. More research is needed to understand the socio-demographic and -economic profile of people resisting AVs.